\documentclass[twocolumn,showpacs,aps,prl,superscriptaddress]{revtex4}

\usepackage{graphicx}
\usepackage{amsmath}

\begin{document}

\title{Modulation of the high mobility two-dimensional electrons in Si/SiGe using atomic-layer-deposited gate dielectric}

\author{K. Lai}
\affiliation{Department of Electrical Engineering, Princeton
University, Princeton, New Jersey 08544}
\author{P.D. Ye}
\affiliation{School of Electrical and Computer Engineering, 
Purdue University, West Lafayette, IN 47907}
\author{W. Pan}
\affiliation{Sandia National Laboratories, Albuquerque, NM
87185}
\author{D.C. Tsui} \affiliation{Department of Electrical
Engineering, Princeton University, Princeton, New Jersey 08544}
\author{S.A. Lyon}
\affiliation{Department of Electrical Engineering, Princeton
University, Princeton, New Jersey 08544}
\author{M. Muhlberger}
\affiliation{Institut fur Halbleiterphysik, Universitat Linz, Linz, Austria}
\author{F. Schaffler}
\affiliation{Institut fur Halbleiterphysik, Universitat Linz, Linz, Austria}

\date{\today}

\begin{abstract}

Metal-oxide-semiconductor field-effect transistors (MOSFET's) using atomic-layer-deposited (ALD) Al$_2$O$_3$ as the gate dielectric are fabricated on the Si/Si$_{1-x}$Ge$_x$ heterostructures. The low-temperature carrier density of a two-dimensional electron system (2DES) in the strained Si quantum well can be controllably tuned from 2.5$\times$10$^{11}$cm$^{-2}$ to 4.5$\times$10$^{11}$cm$^{-2}$, virtually without any gate leakage current. Magnetotransport data show the homogeneous depletion of 2DES under gate biases. The characteristic of vertical modulation using ALD dielectric is shown to be better than that using Schottky barrier or the SiO$_2$ dielectric formed by plasma-enhanced chemical-vapor-deposition(PECVD).

\end{abstract}
\maketitle

Atomic layer deposition (ALD) is a surface controlled layer-by-layer process for the deposition of thin films with atomic layer accuracy. A variety of ALD oxides, such as Al$_2$O$_3$, has been intensively studied as high-k gate dielectrics for microelectronic device applications\cite{wilk}. Similar to SiO$_2$ and Si$_3$N$_4$, Al$_2$O$_3$ can significantly reduce the gate leakage current of metal-oxide-semiconductor field-effect-transistors (MOSFET's). Al$_2$O$_3$ offers additional advantages of a large band gap (9eV), high dielectric constant (k$\sim$8.6-10), high breakdown field (10$^7$V/cm), and thermal stability (amorphous up to at least 1000$^o$C). Furthermore, it can be easily removed by wet-etching and is robust against interfacial reactions and moisture absorption. On the other hand, the research of ALD Al$_2$O$_3$ as the gate dielectric on high mobility two-dimensional electron system (2DES) has been sparse. A device using the ALD Al$_2$O$_3$ as gate dielectric, taking full advantage of its high permittivity, low defect density, high uniformity, conformal step coverage and moderate growth conditions, can be expected to cause least degradation on the quality of the 2DES and provide physical conditions to study yet unexplored low-temperature 2D electronic phenomena. Therefore, work combining ALD and high mobility 2DES would fill the existing gap in ALD applications between basic physics research and state-of-the-art microelectronic device research.

Among the various high mobility 2DES's, the system of electrons confined to the strained Si quantum well in the Si/SiGe heterostructure has emerged as a promising system for the study of 2D electron physics due to its increasing sample quality\cite{schaffler}. Very high electron mobility ($\sim$500,000 cm$^2$/Vs) was reported in this system and strong electron-electron interaction physics phenomena, such as the fractional quantum Hall effect (FQHE)\cite{dunford,laiprl} and the 2D metal-insulator transition (MIT)\cite{olshanetsky, laiapl, laicondmat}, have been observed. However, the vertical modulation on high mobility n-type Si/SiGe proves to be a formidable task due to the difficulty of fabricating a functional field-effect transistor on the heterostructure material. Schottky gate suffers large gate leakage current and only a limited density tuning range can be achieved\cite{ismail, dunford2}. At first glance, SiO$_2$ is the best choice for the gate dielectric. Unfortunately, conventional high temperature oxidation is not amenable to grow SiO$_2$ on modulation doped heterostructures, which are grown at much lower temperatures. Low temperature plasma-enhanced chemical-vapor-deposition (PECVD) grown SiO$_2$ has relative poor quality, i.e., a large amount of traps in bulk SiO$_2$ and at the SiO$_2$/Si interface, compared to thermally oxidized SiO$_2$. In a previous study, a layer of PECVD oxide with thickness less than 100nm cannot prevent a leakage path from the gate to the 2D plane\cite{laiapl}. The low quality of this dielectric, mainly due to the large amount of charged traps inside the SiO$_2$, results in a slow response of the 2DES to the gate bias. As a result, a new gate dielectric material with better electrical properties is needed to modulate the high mobility 2DES in the strained Si quantum well.

In this letter, we report a novel Si/SiGe FET structure using 100nm ALD Al$_2$O$_3$ as the gate dielectric. The low-temperature magnetotransport data show that the 2D carrier density $n$ can be homogeneously depleted from $\sim4.0\times10^{11}$cm$^{-2}$ to below $2.5\times10^{11}$cm$^{-2}$  with a few volts of gate bias (V$_G$) and virtually zero gate leakage current (I$_{leak}$). The observed instantaneous response of the 2DES to the change of V$_G$ at T = 0.3K and a linear $n$ vs. V$_G$ relation manifests the high quality of the ALD dielectric. Results from similar FET structures using either Pd Schottky gate or the PECVD SiO$_2$ as the gate oxide are listed for comparison and the advantage of employing ALD Al$_2$O$_3$ as a gate dielectric is demonstrated.

\begin{figure}[!t]
\begin{center}
\includegraphics[width=2.4in,trim=0.1in 0.6in 0.5in 0.3in]{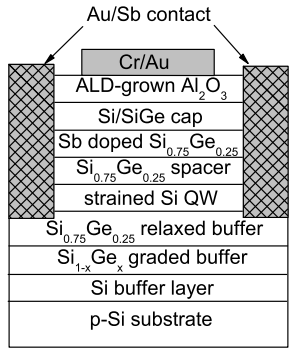}
\end{center}
\caption{\label{1} Schematic view of an n-channel Si/SiGe MOSFET with ALD-grown Al$_2$O$_3$ as gate dielectric. The thickness of the oxide layer is 100nm.}
\end{figure}

Figure 1 shows the device structure of the fabricated n-Si/SiGe MOSFET. The starting material (with an as-grown electron density $n$=4.4$\times$10$^{11}$cm$^{-2}$ and mobility $\mu$=8.0$\times$10$^4$cm$^2$/Vs) is a modulation doped n-type Si/Si$_{1-x}$Ge$_x$ heterostructure grown by molecular-beam epitaxy(MBE). On top of the 100nm Si buffer grown on a p-Si substrate, the relaxed SiGe buffer is realized by a graded layer of 0.5$\mu$m Si$_{1-x}$Ge$_x$ with the Ge mole fraction $x$ varying from 0 to 0.27. A layer of 2.5$\mu$m Si$_{0.75}$Ge$_{0.25}$ is then grown, followed by a 15nm strained Si channel, a 12nm intrinsic Si$_{0.75}$Ge$_{0.25}$ spacer, a 20nm doping layer, a 45nm Si$_{0.75}$Ge$_{0.25}$, and a 10nm Si cap layer. After etched into a 100$\mu$m$\times$320$\mu$m Hall bar, the sample was transferred $ex$ $situ$ to an ASM Pulsar2000$^{TM}$ ALD module. A 100nm-thick Al$_2$O$_3$ layer was deposited at a substrate temperature of 300$^o$C. The oxide on the contact regions was removed by diluted HF and Ohmic contacts were formed by thermal evaporation of 1\%Sb doped Au and 370$^o$C anneal in a forming-gas ambient. The front gate was then formed by thermal evaporation of a Cr/Au layer and lift-off process. After FET fabrication, the sample density and mobility were degraded to $n$=3.9$\times$10$^{11}$cm$^{-2}$ and $\mu$=5.7$\times$10$^4$cm$^2$/Vs at zero gate bias. The sample was mounted in a pumped He$^3$ refrigerator at the base temperature of T = 0.3K. Standard low-frequency ($\sim$7Hz) lock-in techniques were employed to measure the diagonal resistivity $\rho_{xx}$ and the Hall resistance $\rho_{xy}$. A low excitation current, 10nA, was used throughout the experiments.

\begin{figure}[!t]
\begin{center}
\includegraphics[width=2.8in,trim=0.2in 0.4in 0.2in 0in]{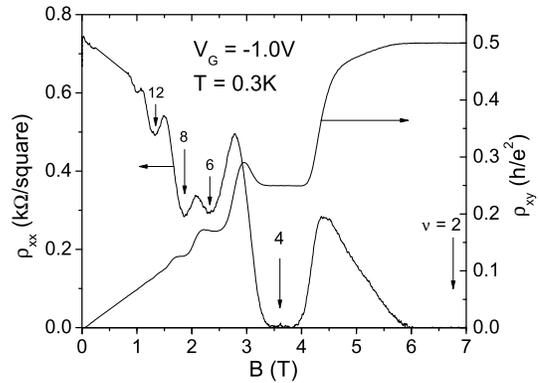}
\end{center}
\caption{\label{2} Longitudinal resistivity $\rho_{xx}$ and Hall resistivity $\rho_{xy}$ as a function of magnetic field at T = 0.3K, with an applied front gate voltage V$_G$ = -1.0V. The electron density and mobility are 3.4$\times$10$^{11}$cm$^{-2}$ and 2.5$\times$10$^4$cm$^2$/Vs, respectively. Major Integer Quantum Hall states are indicated by arrows. The excitation current I$_{ds}$ is 10nA.}
\end{figure}

Figure 2 shows a low temperature magnetotransport trace taken at V$_G$=-1.0V ($n$=3.4$\times$10$^{11}$cm$^{-2}$ and $\mu$=2.5$\times$10$^4$cm$^2$/Vs). The appearance of integer quantum Hall states at Landau level filling factors $\nu$=2, 4, 6, 8 ... as both minima in $\rho_{xx}$ and plateaus in $\rho_{xy}$ demonstrates the high quality of the 2DES under gating. The fact that $\nu$ changes by an even number is due to the remaining two-fold valley degeneracy in the strained Si system, with a valley-valley splitting too small to be resolved here in the presence of disorder broadening.  

\begin{figure}[!t]
\begin{center}
\includegraphics[width=2.8in,trim=0.2in 0.4in 0.2in 0in]{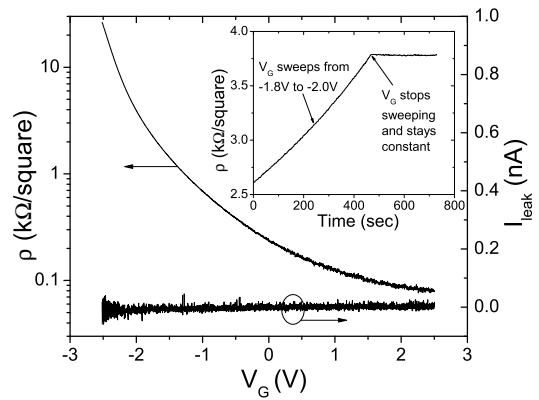}
\end{center}
\caption{\label{3} The four-terminal resistance and the gate leakage current as a function of the gate bias. The inset shows the instantaneous response of 2DES to V$_G$.}
\end{figure}

Figure 3 illustrates the excellent performance of the ALD Al$_2$O$_3$ as a gate dielectric. As the gate voltage is swept from -2.5V to 2.5V, virtually no leakage current is detected and no hysteresis for up and down gate sweeps. However, when the 2DES was depleted into the insulating regime (for V$<$-2.5V and $\rho$$\ge$h/e$^2$), hysteresis in gate sweep was seen, presumably due to the breakdown of screening. More remarkably, as shown in the inset of Figure 3, the 2DES responses to the applied V$_G$ instantaneously and $\rho$ stays constant as the gate bias stops sweeping, in sharp contrast to the case with low-quality PECVD SiO$_2$ as the gate dielectric\cite{laiapl}. 

\begin{figure}[!t]
\begin{center}
\includegraphics[width=2.2in,trim=0in 0.4in 0in 0in]{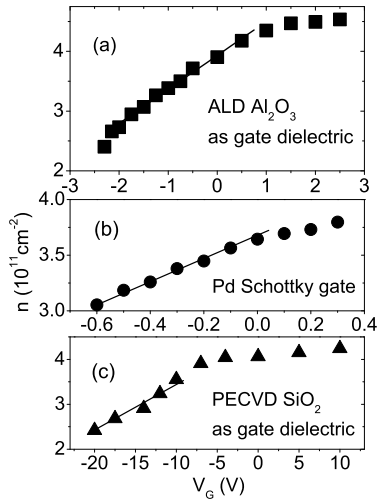}
\end{center}
\caption{\label{4}  Carrier density $n$ vs. V$_G$ for (a) the 100nm ALD Al$_2$O$_3$ gate, (b) the Pd Schottky gate, and (c) the 150nm PECVD SiO$_2$ gate. Straight lines are linear fits to the measured data in the depletion region.}
\end{figure}

To further compare the property of ALD dielectric to that of other gating materials on Si/SiGe, samples with either a Pd Schottky gate or a PECVD SiO$_2$ gate dielectric (deposited at 250$^o$C) were fabricated. Figure 4 shows the $n$ vs. V$_G$ relation for the three different gating schemes, in which the carrier densities are determined by both the quantum oscillations of $\rho_{xx}$ and the low field Hall slopes at 0.3K. As shown in Figure 4a, the carrier density in the ALD sample can be depleted from 4.1$\times$10$^{11}$cm$^{-2}$ to below 2.5$\times$10$^{11}$cm$^{-2}$ as a linear function of the applied gate voltage. A slope as high as 0.6$\times$10$^{11}$cm$^{-2}$/V is extracted in the linear region, even though the device had not received special surface cleaning before the ALD growth or post-growth oxide annealing as in the standard ALD MOSFET process\cite{ye}. Further experiments are needed in order to understand the interface property between the Si and the ALD Al$_2$O$_3$, and it is expected that devices with improved interfaces will bring the capacitance closer to the slope expected of an ideal parallel-plate capacitor -- in this case, $\sim$2.8$\times$10$^{11}$cm$^{-2}$/V. When the device operates in the accumulation regime with V$_G>$ 0V, the data deviate from the linear $n$ vs. V$_G$ relation observed for V$_G$$\le$0V. This deviation is probably due to the accumulation of charges in the Si cap layer. For the Pd Schottky gate (Figure 4b), a gate leakage current in the order of 100pA is detected when V$_G$$<$-0.7V and the density can only be reduced to about 3.0$\times$10$^{11}$cm$^{-2}$. The data shown in Figure 4c were taken from a similarly fabricated device structure, where a 150nm PECVD SiO$_2$ layer was deposited as the gate dielectric instead of 100nm ALD Al$_2$O$_3$. Although the gate leakage current is negligible and the 2D density modulation could also be achieved, the gate oxide shows an inferior characteristic compared to the ALD Al$_2$O$_3$. First, a gate bias voltage as high as $\pm$20V is needed to tune the similar density range, presumably due to the thicker gate oxide, the lower dielectric constant for SiO$_2$ ($\epsilon_{SiO2}/\epsilon_{Al2O3}$=0.39) and much higher trap density in bulk SiO$_2$ and at the interface. Second, the $n$ vs. V$_G$ relation is highly nonlinear even in the depletion side, especially for -10V$<$V$_G$$<$0V. A possible origin of this nonlinearity in gate performance is the traps inside the PECVD oxide and at the interface, which screens the electric field from the metallic gate to the 2D plane.

In summary, we have demonstrated a working field-effect transistor using ALD Al$_2$O$_3$ as gate dielectric on a Si/SiGe heterostructure. The electron density in the strained Si quantum well can be tuned linearly and instantaneously by applying a gate voltage. Neither leakage current nor hysteresis for up and down gate sweeps was observed. Similar FET structures using the Schottky barrier or the gate dielectric formed by PECVD SiO$_2$ are also discussed for comparison and the ALD sample shows the best performance. The experimental technique described above opens up a way to implement ALD oxide as gate dielectrics on high mobility modulation doped heterostructures and to explore 2D physics in previously inaccessible regimes.

We thank J.J. Wang and G.D. Wilk for technical support on ALD growth, and K. Yao and X. Bo for PECVD SiO$_2$ growth. The work at Princeton was supported by AFOSR under grant No. 0190GFB463 and the NSF DMR0352533. 

a) E-mail: klai@princeton.edu   

b) E-mail: yep@purdue.edu

\end{document}